\documentclass[%
 aip,
 amsmath,amssymb,
 reprint,%
]{revtex4-1}

\usepackage{graphicx}
\usepackage{dcolumn}
\usepackage{bm}

\usepackage[utf8]{inputenc}
\usepackage[T1]{fontenc}
\usepackage{mathptmx}
\usepackage{etoolbox}

\makeatletter
\def\@email#1#2{%
 \endgroup
 \patchcmd{\titleblock@produce}
  {\frontmatter@RRAPformat}
  {\frontmatter@RRAPformat{\produce@RRAP{*#1\href{mailto:#2}{#2}}}\frontmatter@RRAPformat}
  {}{}
}%
\makeatother

\usepackage{bm, amsmath, upgreek}

\newcommand{\dd}{\,\mathrm{d}}
\newcommand{\Id}{\mathbb{I}}
\newcommand{\brac}[1]{\left[ #1 \right]^\pm}
\newcommand{\curl}[1]{\left\lbrace #1 \right\rbrace^\pm}
\newcommand{\first}[1]{\tilde{#1}}

\usepackage{xcolor}
\DeclareMathOperator{\spn}{span}

\graphicspath{ {Figures/} }

\begin{document}

\preprint{AIP/123-QED}

\title[Normal mode analysis of fluid discontinuities]{Normal mode analysis of fluid discontinuities:\\numerical method and application to magnetohydrodynamics}
\author{William Béthune}
\email{william.bethune@onera.fr}
\affiliation{DAAA, ONERA, Université Paris Saclay, F-92322 Ch\^atillon, France}

\date{\today}

\begin{abstract}
  Fluid discontinuities, such as shock fronts and vortex sheets, can reflect waves and become unstable to corrugation. Analytical calculations of these phenomena are tractable in the simplest cases only, while their numerical simulations are biased by truncation errors inherent to discretization schemes. The author lays down a computational framework to study the coupling of normal modes (plane linear waves) through discontinuities satisfying arbitrary conservation laws, as is relevant to a variety of fluid mechanical problems. A systematic method is provided to solve these problems numerically, along with a series of validation cases. As a demonstration, it is applied to magnetohydrodynamic shocks and shear layers to exactly recover their linear stability properties. The straightforward inclusion of nonideal (dispersive, dissipative) effects notably opens a route to investigate how these phenomena are altered in weakly ionized plasmas. 
\end{abstract}

\maketitle

\section{Introduction}

Liquids and gases are commonly treated as continuous media, but this assumption becomes unnecessary (or even detrimental) in various circumstances. For example, the thickness of a shock wave $\sim 10^{-5} \mathrm{m}$ in the air \citep{vonmises50,puckett50} and that of liquid-vapor water diffusion layers $\sim 10^{-9} \mathrm{m}$,\citep{matsumoto88,caupin05} which are controlled by small but nonvanishing diffusive effects, are admittedly irrelevant to neighboring meter-scale phenomena. In such cases, one may favor a discontinuous description of the fluid by treating its surfaces of sharp variations as time-varying boundaries. A number of phenomena are of interest in this context, from the reflection and transmission of waves at interfaces to instabilities of the interface itself.

One can study such problems via multidimensional and time-dependent simulations after discretizing the underlying equations, using Riemann solvers for example.\citep{leveque02,toro09} However, such simulations are inherently flawed at their discretization scales and technically limited in the scale separation they can represent. In practice, the internal structure of realistic transition layers is seldom resolved. To tackle the discontinuous limit, analytical methods are laborious if not impossible to carry, in general, especially as research interest concentrate on nonlinear dynamics.\citep[][]{blokhin02} In between, numerical tools often intervene at the end of \emph{ad hoc} semi-analytical calculations, but with no systematically reliable resolution procedure across different problems. I put forward such a procedure applicable to arbitrary systems of conservation laws. 

I recall in Sec. \ref{sec:theory} the jump conditions associated with structurally stable discontinuities in solutions of conservation laws. I take the equations of magnetohydrodynamics (MHD) as an example and perturb their solutions with normal linear modes. I describe in Sec. \ref{sec:method} a procedure to compute the response of arbitrary discontinuities to such perturbations, including the growth rate of interface instabilities. Finally, I provide in Sec. \ref{sec:results} several applications to MHD flows against which analytical and discretized calculations have had a limited reach. In addition to providing a general framework for hyperbolic problems, original results are produced by exactly accounting for nonideal (e.g., parabolic) terms.


\section{Theoretical framework} \label{sec:theory}

\subsection{Conservative equations of magnetohydrodynamics}

Fluid flows are commonly described by the distributions of their primitive variables: the mass density $\rho$, velocity $\bm{V}$, and pressure $P$. Under the assumptions of MHD,\citep{landau84,bellan06} electrically conducting fluids evolve together with the magnetic field $\bm{B}$ they are immersed in. For inviscid and calorically ideal fluids, and after absorbing constant factors into electromagnetic units, the combined Euler's and Maxwell's equations read
\begin{align}
  &\frac{\partial \rho}{\partial t} + \bm{V}\cdot\nabla \rho = -\rho\nabla\cdot\bm{V}, \label{eqn:drho_dt}\\
  &\frac{\partial \bm{V}}{\partial t} + \bm{V}\cdot\nabla\bm{V} = -\frac{\nabla P}{\rho} + \frac{\bm{J}\times\bm{B}}{\rho}, \label{eqn:dV_dt}\\
  &\frac{\partial P}{\partial t} + \bm{V}\cdot\nabla P = -\gamma P \nabla\cdot\bm{V} + \left(\gamma-1\right) \bm{E} \cdot \bm{J}, \label{eqn:dP_dt}\\
  &\frac{\partial \bm{B}}{\partial t} + \bm{V}\cdot\nabla\bm{B} = -\bm{B} \nabla\cdot\bm{V} +\bm{B}\cdot\nabla\bm{V} -\nabla\times \bm{E}, \label{eqn:dB_dt}
\end{align}
where $\bm{J} = \nabla\times\bm{B}$ is the electric current density in the nonrelativistic limit, $\nabla\cdot\bm{B}=0$ following Gauss's law, and $\gamma$ is the adiabatic index appearing in the equation of state $P = (\gamma-1)\rho \epsilon$ given a specific internal energy $\epsilon$. One can otherwise prescribe the sound speed $c_s$ in the isothermal equation of state $P=\rho c_s^2$ and discard energy considerations altogether.

For imperfect electrical conductors, the electric field $\bm{E}$ need not vanish in the frame comoving with the fluid. Let us incorporate Ohmic resistivity, the Hall drift, and ambipolar diffusion in the following expansion appropriate for collisional and weakly ionized plasmas: \citep[e.g.,][]{pandey08}
\begin{equation} \label{eqn:emf}
  \bm{E} = \eta \bm{J} + \frac{\lambda}{\sqrt{\rho}} \bm{J}\times \bm{B} - \frac{\tau}{\rho} \left(\bm{J}\times\bm{B}\right)\times\bm{B}.
\end{equation}

Alternatively, one can define the following set of conservative variables: the mass density $\rho$, momentum density $\bm{m} = \rho \bm{V}$, magnetic field $\bm{B}$, and total energy density $\mathcal{E} = \rho \epsilon + (\rho V^2 + B^2)/2$. Every conservative variable $q$ obeys an equation of the form
\begin{equation} \label{eqn:dq_dt}
  \frac{\partial q}{\partial t} + \nabla \cdot \bm{f}_q = s_q, 
\end{equation}
where $\bm{f}_q$ is the flux of $q$ and $s_q$ accounts for possible source terms, such as accelerations in noninertial frames. Combining the previous equations leads to the following expressions for the mass flux vector $\bm{f}_{\rho} = \rho \bm{V}$, the momentum flux tensor $\bm{f}_{\bm{m}} = \rho \bm{V}\otimes\bm{V} + (P + B^2/2) \Id - \bm{B}\otimes\bm{B}$, where $\Id$ is the identity matrix, and the energy flux vector $\bm{f}_{\mathcal{E}} = (\rho \epsilon + \rho V^2/2 + P)\bm{V} + \bm{E}^\prime\times \bm{B}$, where $\bm{E}^\prime = \bm{E} -\bm{V}\times\bm{B}$. The flux associated with the magnetic field can be written $\bm{f}_{\bm{B}} = \bm{V}\otimes\bm{B} - \bm{B}\otimes\bm{V}$ in ideal MHD, whereas the nonideal effects in Eq. \eqref{eqn:emf} introduce additional spatial derivatives in the flux via the electric current density. As we shall see, the method presented below relies on the conservative form of Eq. \eqref{eqn:dq_dt}.

\subsection{Rankine-Hugoniot relations}

It is known that inviscid flows admit solutions with arbitrarily steep gradients and that Euler's equation allows the formation of discontinuities out of smooth initial conditions in finite times (e.g., via wave steepening). Let us consider structurally stable discontinuities as smooth interfaces such that, for any point on the interface, one can define a unit normal vector $\bm{n}$ and locally reduce Eq. \eqref{eqn:dq_dt} to the normal direction: $\partial_t q + \partial_z (\bm{f}_q\cdot \bm{n})=s_q$, where $z=\bm{x}\cdot\bm{n}$. Integrating the first term on a neighborhood of the interface at $z=\zeta(x,y,t)$ yields
\begin{align} \label{eqn:weak}
  \begin{split}
    \int_{\zeta-\delta}^{\zeta+\delta} \frac{\partial q}{\partial t} \dd z &= \frac{\partial}{\partial t} \int_{\zeta-\delta}^{\zeta} q \dd z + \frac{\partial}{\partial t} \int_{\zeta}^{\zeta+\delta} q \dd z \\
    &-\left[q\left(\zeta+\delta\right)-q\left(\zeta-\delta\right)\right] \frac{\partial \zeta}{\partial t}.
  \end{split}
\end{align}
Similarly integrating the other terms, assuming that source terms remain finite, and taking the limit $\delta \rightarrow 0$ yields
\begin{equation} \label{eqn:RH}
  \brac{q} \frac{\partial \zeta}{\partial t} - \brac{\bm{f}_q} \cdot \bm{n} = 0, 
\end{equation}
where $\brac{X} = X^+ - X^-$ denotes the difference between the values $X(\zeta+\delta)$ and $X(\zeta-\delta)$ on both sides of the interface as $\delta \rightarrow 0$. Equation \eqref{eqn:RH} contains the jump conditions that must be satisfied by all conservative variables at the interface, and is known as the Rankine-Hugoniot relation. In the following, I will always adopt steady initial conditions ($\partial_t\zeta=0$) whose normal fluxes are, therefore, continuous. 

\subsection{Free normal modes}

Let us decompose every quantity on each side of a given discontinuity as the sum of a constant background value plus fluctuations: $X = \overline{X} + \first{X}$, so that $\partial_t \first{q} + \nabla\cdot \first{\bm{f}_q} = \first{s}_q$. Gathering the system of equations for the first-order perturbations of the conservative variables gives
\begin{equation} \label{eqn:lindq_dt}
  \frac{\partial \first{\bm{Q}}}{\partial t} + \nabla \cdot \left(\frac{\partial \bm{F}}{\partial \bm{Q}}\bigg\rvert_{\overline{\bm{Q}}} \cdot \first{\bm{Q}}\right) - \frac{\partial \bm{S}}{\partial \bm{Q}}\bigg\rvert_{\overline{\bm{Q}}} \cdot\first{\bm{Q}} = 0,
\end{equation}
where it is assumed that the matrices $\partial \bm{F} / \partial \bm{Q}$ have only real eigenvalues in the absence of dissipative effects: the media support waves, they are not unstable on their own. 

One then needs to deal with the possibly complicated geometry of wave fronts and of the interface itself. To circumvent most difficulties, I focus on normal modes $\sim \exp (i [\omega t - \bm{k} \cdot \bm{x}])$ with complex frequencies $\omega$ and $\bm{k}$, whose period $2\uppi/\vert\omega\vert$ is much shorter than the global evolution timescale of the interface, and whose wavelength $2\uppi/\|\bm{k}\|$ is much smaller than the local curvature radius of the interface. Accordingly, the interface can be seen as a nearly flat and steady plane. In this WKBJ approximation, Eq. \eqref{eqn:lindq_dt} takes the form of an eigenvalue problem, $(\mathcal{A}(\bm{k}) - \omega\Id)\cdot\first{\bm{Q}}=0$ for $(\omega,\first{\bm{Q}})$ given $\bm{k}$,
and whose characteristic polynomial is known as the dispersion relation, $\mathcal{D}(\omega,\bm{k})=0$. An eigenmode is said to be unstable if the corresponding eigenvalue has $\Im(\omega) < 0$.

\subsection{Mode coupling and statement of the problem}

Let us keep a Cartesian coordinate system $(x,y,z)$ such that the normal to the interface is initially along $z$. Under the previous assumptions, the problem loses explicit dependencies on time and on the coordinates $(x,y)$ tangent to the interface. Because of these symmetries, a set of plane waves on both sides of the interface cannot satisfy the jump conditions \eqref{eqn:RH} for all $(x,y,t)$ unless they share the same $(k_x,k_y,\omega)$.

It also becomes apparent that the perturbed locus of the interface $\first{\zeta}(x,y,t)$ varies exponentially with the same arguments $(k_x,k_y,\omega)$. To first order in perturbations, the unit normal vector becomes $\bm{n} = \overline{\bm{n}} + \first{\bm{n}}$, with $\overline{\bm{n}}=(0,0,1)$ and $\first{\bm{n}} = (-\partial_x\first{\zeta},-\partial_y\first{\zeta},0)$. Taking into account the fluctuations of $\first{\zeta}$ in the integration bounds [cf. Eq. \eqref{eqn:weak}], the linearized Rankine-Hugoniot relations read
\begin{equation} \label{eqn:linRH}
  \brac{\overline{q}} \frac{\partial \first{\zeta}}{\partial t} - \brac{\first{\bm{f}}_q} \cdot \overline{\bm{n}} - \brac{\overline{\bm{f}}_q} \cdot \first{\bm{n}} + \brac{\overline{s}_q}\first{\zeta} = 0. 
\end{equation}

The problem is to identify a set of modes on both sides of an interface such that each mode satisfies their respective governing equation \eqref{eqn:lindq_dt}, and there exists a combination of them that satisfies the Rankine-Hugoniot relations \eqref{eqn:linRH} as a whole. 


\section{Method} \label{sec:method}

I now describe a systematic method to solve the problem above using common numerical techniques. 

\subsection{Steady discontinuities}

The first step is to exhibit a discontinuous solution of Eq. \eqref{eqn:dq_dt}. One can always adopt a frame comoving with the interface and reduce it to $\brac{f_q}=0$ for all conservative variables. This root-finding problem is straightforward to solve provided an educated guess for the desired discontinuity. One should only verify that the solution is indeed discontinuous, since the trivial case $\brac{q}=0$ always exists. One should also ensure that the fluxes (e.g., viscous) remain finite at the interface, or else a discontinuity should instantaneously be smeared.

Source terms vanish at this stage for conservative systems, although they appear for first-order perturbations in Eq. \eqref{eqn:linRH}. In fact, the relation $\brac{f_q}=0$ refers to the fluxes at the interface, while source terms would typically induce flux gradients away from the interface. Although the WKBJ approximation is tailored to deal with slowly varying backgrounds, such variations would give a global character to the problem. For example, wave propagation may become forbidden outside a bounded spatial domain. To avoid these complications, I focus hereafter on situations with vanishing background sources ($\overline{s}_q=0$). In particular, I exclude nonconservative cases where source terms may become infinite at the interface. 

\subsection{Dispersion relations}

As far as $k_x$ and $k_y$ are constants independent of the side relative to the interface, they can be seen as free parameters such that the dispersion relation $\mathcal{D}(\omega, k_z\,; k_x, k_y)=0$ is foremost an implicit relation between $\omega$ and $k_z$. For a stability analysis, one would typically prescribe $k_z$ and ask whether there is a corresponding $\omega$ with a negative imaginary part. Instead, let us prescribe the complex frequency $\omega_0$ and seek the set of $k_z$ satisfying $\mathcal{D}(\omega_0,k_z)=0$.

The dispersion relation resulting from a normal mode expansion is generally a multivariate polynomial. To construct it, one can use the recursive cofactor expansion of the matrix $(\mathcal{A}(k_z) - \omega_0\Id)$ after observing that its entries are themselves polynomials in $k_z$. Matrix factorization (e.g., LUP) is ill-advised here due to rounding errors on polynomial coefficients, and to the absence of numerically safe pivot for polynomial division. One can then use polynomial root-finding techniques (e.g., via the companion matrix) to compute the set of $k_z$ satisfying $\mathcal{D}(\omega_0,k_z)=0$ given $\omega_0$. The degree of this characteristic polynomial can be large, favoring errors on the estimated roots that affect negatively the rest of the method. I found it beneficial to iterate Halley's method \citep{scavo95} a few times after multiple roots have been singled-out and to deflate the polynomial by its accurate roots in the most difficult cases.

For each root $k_z$, one can inject the corresponding normal mode into Eq. \eqref{eqn:lindq_dt}, solve the resulting eigenvalue problem for $(\omega, \first{\bm{Q}})$, and store the eigenmodes having $\omega=\omega_0$. Repeating this procedure on both sides of the interface, one ends up with $L$ modes on the left (minus) side and $R$ modes on the right (plus) side that share the same prescribed $(k_x,k_y,\omega)$ and are, therefore, susceptible to match at the interface. This root-finding step is the main hindrance to symbolic calculations. 

\subsection{Compatibility conditions} \label{sec:compatibility}

The following step is to enforce the linearized Rankine-Hugoniot relations onto the available set of normal modes. Replacing the partial derivatives in Eq. \eqref{eqn:linRH}, one obtains
\begin{equation} \label{eqn:compat}
  \first{f}_{q,z}^+ - \first{f}_{q,z}^- - i \left(\omega \brac{\overline{q}} - k_x \brac{\overline{f}_{q,x}} - k_y \brac{ \overline{f}_{q,y}} \right) \first{\zeta} = 0.
\end{equation}
Putting together the relations for all conservative variables produces a linear system of the form $\mathcal{H}\cdot(\first{\bm{F}}_z^-, \first{\bm{F}}_z^+, \first{\zeta})^T=0$, meaning that the admissible perturbations of $(\first{\bm{F}}_z^-, \first{\bm{F}}_z^+, \first{\zeta})$ are in the nullspace (kernel) of $\mathcal{H}$.

Given the jumps in background states, one can explicitly construct $\mathcal{H}$ and a basis for its nullspace. If there are $N$ conservative equations, then $\mathcal{H}$ has dimensions $N\times(2N+1)$, and its nullspace has dimension $N+1$. Indeed, taking $\first{\zeta}=0$ yields $N$ independent vectors that have $\first{f}_{q,z}^-=\first{f}_{q,z}^+=1$ for a single variable $q$. Taking $\first{\zeta}=1$ and $\first{f}_{q,z}^-=0$ for all $q$ yields the last nullspace dimension trivially. The coordinate $\first{\zeta}$ can then be discarded without consequences in the following.

Since Eq. \eqref{eqn:lindq_dt} is typically solved for the perturbed conservative variables $\first{\bm{Q}}^\pm$, one can multiply them by their respective Jacobian matrix $\partial \bm{F}_z / \partial \bm{Q} \rvert_{\overline{\bm{Q}}}$ to obtain the corresponding flux perturbations $\first{\bm{F}}_z^\pm$. This choice of variables is necessary to account for nonideal (e.g., parabolic) effects, when $\omega(k)$ is nonlinear and, hence, conservative fluxes depend on wavelength. Otherwise, one may instead express Eq. \eqref{eqn:compat} in terms of $(\first{\bm{Q}}^-, \first{\bm{Q}}^+)$ fluctuations via the same Jacobian matrices. 

Let us denote the left modes by $\bm{\ell}_i$, the right modes by $\bm{r}_i$, and the basis vectors of the nullspace of $\mathcal{H}$, deprived of the $\first{\zeta}$ coordinate, by $\bm{h}_i$. Imposing the Rankine-Hugoniot relations corresponds to finding linear combinations of $\bm{\ell}$ and $\bm{r}$ vectors that fall in the nullspace of $\mathcal{H}$. This is equivalent to finding the nullspace of $\mathcal{M} = (\bm{\ell}_1,...,\bm{\ell}_L,\bm{r}_1,...,\bm{r}_R,\bm{h}_1,...,\bm{h}_{N+1})$. Any vector in this nullspace describes a superposition of modes that satisfies Rankine-Hugoniot, as encoded in its first $L+R$ components.

\subsection{Regularity conditions} \label{sec:regularity}

It is possible to incorporate nonideal (dissipative, dispersive) effects at the cost of additional regularity constraints. For example, velocity discontinuities cannot subsist in the presence of viscosity and must be precluded from the outset. Otherwise, one could construct discontinuous solutions whose left and right fluxes satisfy the Rankine-Hugoniot relations \eqref{eqn:RH} while the flux evaluated at the interface is actually infinite.

Regarding magnetic fields, their associated fluxes are the components of the electromotive force $\bm{E}$. Both the Ohmic, Hall, and ambipolar terms are proportional to the electric current density $\bm{J} = \nabla\times\bm{B}$, so singularities can arise from linear combinations of the form $E = \sum_i a_i(z) \partial_z B_i$. The requirement of a finite flux translates into $\int_{\zeta-\delta}^{\zeta+\delta} E \dd z \rightarrow 0$ as $\delta\rightarrow 0$. Writing $B$ as a continuous function plus a (Heaviside) step and symmetrically mollifying it, the previous constraint becomes $\sum_i (1/2)(a_i^+ + a_i^-) (B_i^+ - B_i^-)=0$. Such constraints can be implemented via additional rows in the matrix $\mathcal{M}$. It suffices to append each vector $(\bm{\ell},\bm{r})$ with its contribution $\pm \sum_i ( a_i^+ + a_i^-) B_i^\pm$ to the integral of $E$, and to pad the $\bm{h}$ vectors with as many zeros as the number of constraints.

\subsection{Types of solutions} \label{sec:types}

Different situations can arise depending on the rank of $\mathcal{M}$, or implicitly on the imposed value of $\omega$. 

In the first case, $\mathcal{M}$ has a nontrivial nullspace by construction so our problem immediately admits solutions. This necessarily occurs when the number of available modes $L+R \geq N$. The latter condition is typically satisfied when the prescribed $\omega$ is real such that both sides of the interface support enough in and outgoing waves. In this situation, it may be possible to isolate one incident wave at a time, based on the sign of $\omega/k_z$, as if it was forcing the system. A basis of the nullspace of $\mathcal{M}$ can be chosen to reflect this property by canceling the amplitudes of all but one incident wave per basis vector, thus providing the reflection and transmission coefficients onto all the other wave modes.

In the second case, one may seek solutions localized near the interface by restricting the set of interacting modes to those having $\Im({k_z^-})>0$ and $\Im({k_z^+})<0$ and find that $\ker(\mathcal{M})=\{\bm{0}\}$ in general. Only specific values of $\omega \in \mathbb{C}$ allow solutions to be found for given $(k_x,k_y)$ and, hence, the temporal and spatial frequencies, $\omega$ and $k_z$, must simultaneously be solved for. One way to achieve this is to iteratively minimize the smallest singular value of $\mathcal{M}$ as a function of $\omega$, aiming for the minimum to reach zero. An optimal eigenvalue $\omega$ need not be associated with more than one eigenmode on each side of the interface ($L+R<N$). Symmetry considerations can guide the choice of a starting guess for $\omega$, as well as help to find conjugate solution branches.

\subsection{Validation cases}

I gathered below a series of exact results helpful in testing an implementation of the above method, and I walk through one example in Appendix \ref{app:example}. Having recovered all of them with over nine digits of accuracy, comparative illustrations seem superfluous. Let us define $\curl{X} = X^+ + X^-$ in addition to $\brac{X} = X^+ - X^-$, referring here to steady state variables.

\subsubsection{Contact discontinuities}

Contact discontinuities have $\bm{V}=0$ but $\brac{\rho}\neq 0$ and, hence, different sound speeds $c_s=\sqrt{\gamma P/\rho}$ in the absence of magnetic fields. Defining $k_t^2 = k_x^2+k_y^2$, the reflection coefficient for sound waves is given by $\brac{\rho a} / \curl{\rho a}$, where $a=k_t/k_z$ for oblique wave incidence ($k_t\neq 0$), or $a=c_s$ for otherwise normal wave incidence.\citep{landau87}

Contact discontinuities are subject to the Rayleigh-Taylor instability when accelerated toward the high-density side, that is, under a normal momentum source term $s_{m_z}=\rho g$. The resulting pressure stratification is often disregarded by virtue of incompressibility. One can approach this incompressible limit when $g/k c_s^2\ll 1$ and, thus, recover the theoretical growth rate $\omega_{\rm{RT}}=-i\sqrt{a g k_t}$, where $a=\brac{\rho}/\curl{\rho}$. \citep{rayleigh1882,taylor1950}

\subsubsection{Tangential discontinuities} \label{sec:tangent}

Tangential discontinuities (vortex sheets) have $\brac{V_x}\neq 0$ but $V_z=\brac{\rho}=0$, while one can enforce $\brac{V_y} = 0$ by an appropriate frame rotation. This infinitely narrow shear layer admits two kinds of behaviors. On the one hand, the interface can reflect and transmit sound waves, with a reflection coefficient given by $\brac{(\omega-\bm{k}\cdot\bm{V})^2 / k_z} / \curl{(\omega-\bm{k}\cdot\bm{V})^2 / k_z}$.\citep{miles57,landau87}

On the other hand, tangential discontinuities are subject to the Kelvin-Helmholtz instability. Its theoretical growth rate $\omega_{\rm{KH}} = k_x \curl{V_x}/2 - k_x \sqrt{(v/2)^2 + c_s^2 - c_s \sqrt{c_s^2 + v^2}}$ for planar modes ($k_y=0$), where $v=\brac{V_x}$. \citep{landau87} This case involves one acoustic mode traveling through the interface, and one can verify that a shear $v/c_s>\sqrt{8}$ stabilizes it. A constant magnetic field $B_x\neq 0$ also stabilizes incompressible perturbations whenever $\curl{B_x^2} > \rho v^2/2$, even in nonplanar cases. \citep{fejer64,ray83,landau84}

\subsubsection{Hydrodynamic shocks}

Hydrodynamic shocks feature a supersonic normal velocity jump $\brac{V_z}\neq 0$, while the tangential velocity can be brought to zero with an appropriate frame translation. The upstream fluid is necessarily supersonic in this comoving frame, implying that all upstream sound waves propagate toward the interface.

Let $M = V_z / c_s$ denote the flow Mach number and $a=1/M^2$. Considering a sound wave incident from the downstream (plus) side, it can only be reflected with relative amplitude $-(1+a^- -2M^+) / (1+a^- +2M^+)$. Inversely, a sound wave incident from the upstream (minus) side can only be transmitted, and the relative amplitude of the downstream pressure perturbation is given by $((1 + M^-)^2 - b ) /  (1+a^- +2M^+)$, where $b=a^- (1/a^- -1)^2 (\gamma-1) /(\gamma+1)$.\citep{landau87} Additional test cases may be designed on the thermodynamic stability and spontaneous emission of sound by shock waves.\citep{dyakov54,kontorovitch58,swan75}

\subsubsection{Nonideal MHD effects}

Nonideal MHD effects alter the dispersion relation of hydromagnetic waves. To test their proper implementation, one can look for transverse perturbations on a constant magnetic field $\bm{B}$, with a wave vector $\bm{k}$ along $\bm{B}$. In ideal MHD, one finds a pair of Alfv\'en waves with velocity $V_A = B / \sqrt{\rho}$. With Ohmic resistivity in $\bm{E} = \eta \bm{J}$, Alfv\'en modes are damped at frequencies $\omega_{\pm} = k V_A (i a \pm \sqrt{1-a^2})$, where $a = \eta k / 2 V_A$. As far as linear perturbations are concerned, ambipolar diffusion is analogous to Ohmic resistivity, albeit anisotropically. This becomes apparent when writing it as $\bm{E} = \eta_a \bm{J}_{\perp}$, where $\bm{J}_{\perp} = ( \Id - \bm{B}\otimes\bm{B} / \bm{B}\cdot\bm{B}) \cdot \bm{J}$ is the electric current density perpendicular to the magnetic field. When it comes to the Hall drift, it splits Alfv\'en waves into dispersive modes with $\omega_{\pm}^2 = k^2 V_A^2(1 + a \pm \sqrt{(1 + a)^2 - 1})$, where $a = \lambda^2 k^2 / 2$.


\section{Application to magnetized flows} \label{sec:results}

MHD flows support Alfv\'en waves that propagate along field lines due to magnetic tension. Acoustic waves also split into slow and fast magnetosonic branches, for which magnetic pressure contributes as a restoring force. Adding these characteristic modes renders the dispersion relation tedious to solve analytically and entails more diverse interactions at interfaces. I apply the method described above to seamlessly account for this complexity and, thus, obtain exact predictions on instabilities of magnetized discontinuities. After touching base with known ideal MHD cases, I produce original results including nonideal conductivity effects. To isolate their role, these effects are included separately while keeping the thermodynamics simple, omitting radiation transport in particular. Grid-based simulations are only briefly discussed in Appendix \ref{app:dns} to highlight the difficulty of these problems.

\subsection{Parallel slow shocks} \label{sec:idealshock}

Shock fronts threaded by a magnetic field can be unstable to corrugation modes.\citep{lessen67,edelman1989b,stone95} In the parallel case where the magnetic field is normal to the interface (only $B_z\neq 0$), instability requires an upstream Alfv\'en velocity greater than the upstream flow velocity so that Alfv\'en and fast-magnetosonic waves can propagate against the flow upstream of the shock. 

\subsubsection{Ideal MHD}

\begin{figure}
  \includegraphics[width=\linewidth]{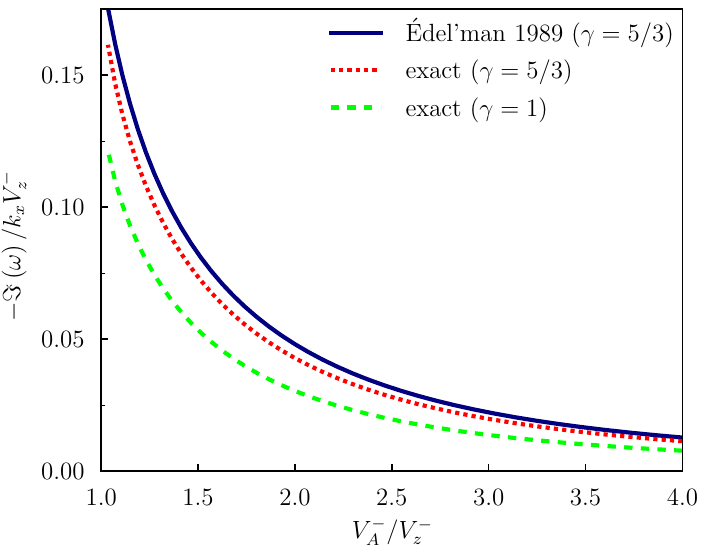}
  \caption{\label{fig:SE95} Corrugation growth rate of parallel slow shocks in ideal MHD. The solid blue line is the asymptotic prediction of {\'E}del'man,\citep{edelman1989b} the dotted red line is the corresponding exact solution, and the dashed green line is the equivalent isothermal case.}
\end{figure}

In ideal MHD, the instability involves one upstream and two downstream magnetosonic modes, plus the downstream entropy-vortex mode for adiabatic shocks. They all have purely imaginary $\omega$ and $k_z$. Figure \ref{fig:SE95} shows the growth rate of the instability as a function of the inverse Alfv\'en number of a shock with upstream Mach number $V_z^- / c_s^- = 2$. The analytical prediction of {\'E}del'Man,\citep{edelman1989b} which was derived in the asymptotic limit $V_z^- / c_s^- \gg 1$ [see their Eq. (29)], is in remarkable agreement with the exact solution. This figure also shows that the instability extends to the isothermal limit, labeled $\gamma=1$, with only slightly reduced growth rates. 

Although the agreement on Fig. \ref{fig:SE95} is expectedly imperfect, it serves as a final validation case of the method. To focus on nonideal MHD effects in the following, let us use an isothermal equation of state $P=\rho c_s^2$, and set the value of the upstream Mach number $V_z^- / c_s^- = 2$.

\subsubsection{Resistive MHD}

\begin{figure}
  \includegraphics[width=\linewidth]{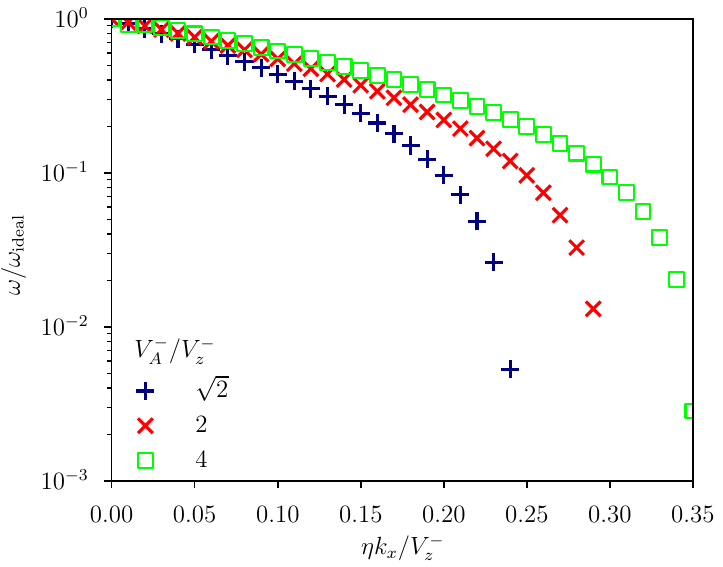} 
  \caption{\label{fig:ohmshock} Corrugation growth rate of parallel slow shocks in resistive MHD relative to the ideal MHD limit. The shock is taken to be isothermal with upstream Mach number $V_z^- / c_s = 2$.}
\end{figure}

As resistivity decouples the magnetic field from the fluid, one should recover the hydrodynamic limit (i.e., stability) at large enough resistivity. Figure \ref{fig:ohmshock} shows how the corrugation growth rate indeed decreases with Ohmic resistivity for three different Alfv\'en numbers of the shock. All three curves feature a cutoff beyond which the growth rate drops to zero, located near $\eta k_x / V_z^- \approx 0.2$, $0.25$, and $0.3$ for $V_A^- / V_z^- = \sqrt{2}$, $2$, and $4$, respectively. Appendix \ref{app:dns} illustrates how challenging it is to recover these results using grid-based simulations. 

One can generally take $k_y=\first{B}_y=0$ in this planar case, implying that the normal current $\first{J}_z$ plays no role on the instability of parallel shocks. Projecting the electric current perpendicular to the magnetic field yields $\first{J}_{\perp}=(\first{J}_x, \first{J}_y, 0)$ to first order in perturbations. Because the $(x,y)$ components are unchanged by projection, ambipolar diffusion is strictly equivalent to Ohmic resistivity during the linear stage of this instability, as far as it is treated in the single-fluid approximation,\citep{wardle91,snow21} with a corresponding diffusivity $\eta_a = \tau \overline{B}_z^2 / \rho$.

\subsubsection{Hall MHD}

\begin{figure}
  \includegraphics[width=\linewidth]{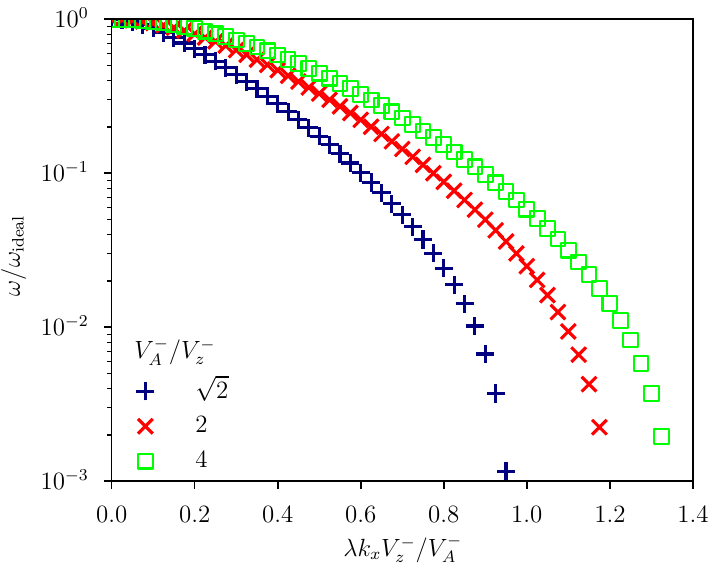}
  \caption{\label{fig:hallshock} Same as Fig. \ref{fig:ohmshock} but in Hall MHD.}
\end{figure}

In addition to altering the frequency of transverse waves, the Hall drift also causes their polarization plane to rotate, making its influence rather nonintuitive. At high spatial frequencies, the ``whistler'' modes propagate at velocities $\sim \lambda k V_A$, so one can always find short-wavelength perturbations propagating against the flow upstream of the shock. As a consequence, all parallel shocks are somewhat slow in Hall MHD, and one may be tempted to extrapolate ideal MHD conclusions regarding their instability. Against such expectations, I show in Fig. \ref{fig:hallshock} that the Hall drift actually weakens the corrugation instability at small scales and kills it beyond some critical $\lambda k_x \sim V_A^- / V_z^-$ in the adopted configuration.

\subsection{Tangential velocity discontinuities}

In this section I consider vortex sheets with $\brac{V_x}\neq 0$ embedded in a constant magnetic field $B_x \neq 0$. I omit electric currents initially and, therefore, exclude issues of magnetic reconnection.\citep{min96,keppens99,faganello17} I take the fluid to be isothermal and examine the influence of nonideal MHD effects on the Kelvin-Helmholtz instability, reminding that strong enough magnetic fields can stabilize it in ideal MHD (see Sec. \ref{sec:tangent}).

\subsubsection{Ohmic resistivity}

\begin{figure}
  \includegraphics[width=\linewidth]{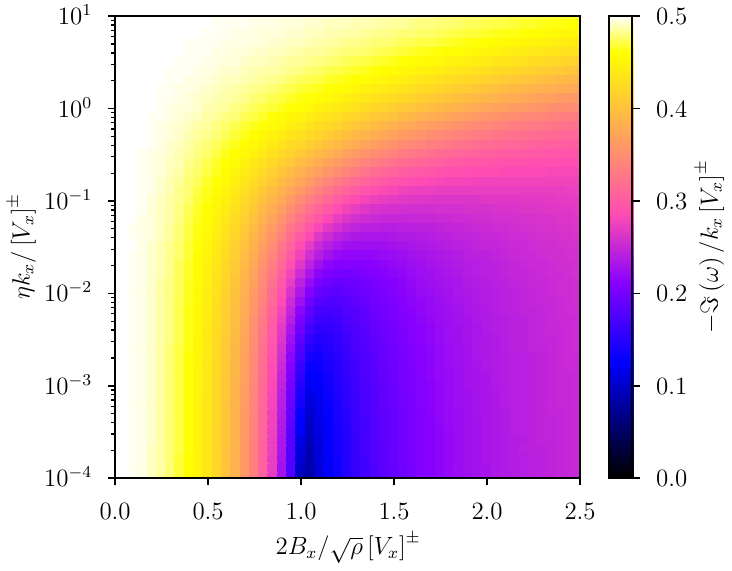}
  \caption{\label{fig:ohmshear} Growth rate of planar Kelvin-Helmholtz modes as a function of magnetic strength and resistivity given a shear $\brac{V_x}=2\times 10^{-2}c_s$.}
\end{figure}

Because Ohmic resistivity acts against the buildup of a stabilizing magnetic tension, it should enhance or even revive the instability in magnetized flows. Figure \ref{fig:ohmshear} shows how the corrugation growth rate varies with resistivity and magnetic field strength in a planar ($k_y=0$) and very subsonic case: $\brac{V_x}=2\times 10^{-2} c_s$. The growth rate tends to $k_x \brac{V_x}/2$ in the hydrodynamic limits of weak field or large resistivity, as expected. In the other limit, a small but nonvanishing resistivity allows fast-growing modes to be found beyond the ideal stability threshold of $B_x/\sqrt{\rho} \simeq \brac{V_x}/2$. These unstable solutions involve modes with $k_z /k_x \gg 1$ in magnitude, leading to an ever greater scale separation as resistivity decreases. One may attribute them to the magnetic Reynolds number $\brac{V_x}/\eta k_z \ll 1$ at sufficiently small scales. 

\subsubsection{The Hall drift}

\begin{figure}
  \includegraphics[width=\linewidth]{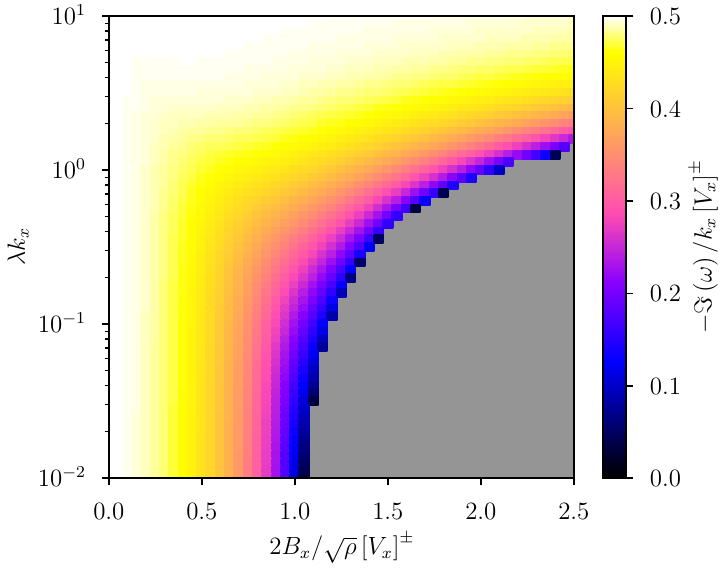}
  \caption{\label{fig:halshear} Same as Fig. \ref{fig:ohmshear} but in Hall MHD. The gray area is stable.}
\end{figure}

The Hall drift is known to destabilize shear flows in various configurations.\citep{kunz08,pandey18} As previously, let us consider planar perturbations over a subsonic shear $\brac{V_x}=2\times 10^{-2} c_s$ and start with a constant magnetic field $B_x$ that is parallel to the flow. The intensity of the Hall drift is tuned via its relative inertial length $\lambda k_x$ after prescribing a tangential wave number $k_x$. 

Figure \ref{fig:halshear} shows how the growth rate of corrugation modes varies with magnetic field strength and Hall inertial length. When $\lambda k_x \ll 1$ (bottom side), one recovers the ideal regime where a magnetic field $B_x/\sqrt{\rho} \gtrsim \brac{V_x}/2$ stabilizes the flow. As $\lambda k_x$ increases, the instability's growth rate increases and the unstable region extends to stronger magnetic fields. In this scenario, the critical field strength is roughly doubled at scales $\lambda k_x \sim 1$. These results refine and extend previous asymptotic predictions into the compressible regime. \citep{pandey18}

\subsubsection{Ambipolar diffusion}

\begin{figure}
  \includegraphics[width=\linewidth]{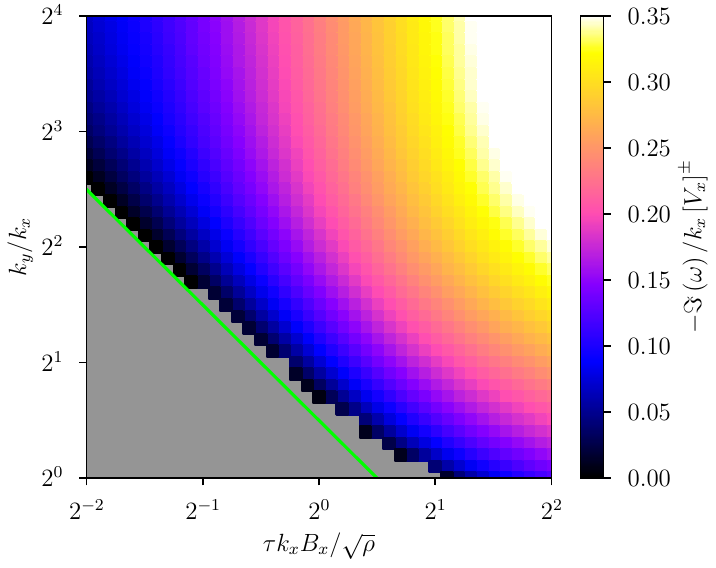}
  \caption{\label{fig:ambshear} Growth rate of Kelvin-Helmholtz modes as a function of ambipolar diffusion time and perturbation obliqueness. The prescribed shear $\brac{V_x}=3 c_s$ and the magnetic field strength $B_x/\sqrt{\rho}=\brac{V_x}$. The gray area below the green line is stable.}
\end{figure}

In ambipolar MHD, the projection of electric currents perpendicular to a uniform $\overline{B}_x$ yields $\first{J}_{\perp}=(0, \first{J}_y, \first{J}_z)$, so the $x$ component of the nonideal electromotive force vanishes. As a consequence, $\first{B}_z$ does not diffuse along $y$ and $\first{B}_y$ does not diffuse along $z$. The action of ambipolar diffusion on oblique modes is, therefore, highly nontrivial. 

Let us take $\brac{V_x} = 3 c_s$ to ensure that planar ($k_y=0$) modes are stable without magnetic fields, and $B_x/\sqrt{\rho}=\brac{V_x}$ to also stabilize nonplanar modes in ideal MHD. Figure \ref{fig:ambshear} shows the growth rates of unstable solutions found over a range of $k_y/k_x$ and ambipolar collision times $\tau$. Instability requires sufficiently large $\tau$ to approach the hydrodynamic regime and large $k_y/k_x$ as expected in this regime, although single-fluid models become inappropriate when $\omega \tau \gg 1$.\citep{watson04} Inspecting the growing modes reveals that they feature a discontinuity in $\first{B}_y$ at the interface, which is specifically allowed by ambipolar diffusion in this case.


\section{Concluding remarks}

I considered discontinuous solutions of conservation laws typical in fluid flows and examined the coupling of small-amplitude fluctuations living on both sides of the interface. Following a normal mode decomposition, I presented a systematic method to compute all the reflection and transmission coefficients at the interface and the growth rate of interface instabilities. After testing an implementation of this method against known exact results, I refined previous predictions on the corrugation instability of parallel slow MHD shocks. I then incorporated nonideal MHD effects to produce original and exact results on instabilities of magnetized but weakly ionized shocks and tangential discontinuities. 

The formalism is purposefully general and may be adapted to include other phenomena, such as interface curvature,\citep{wada04,lee12} surface tension,\citep{levich69,malkin08} or internal energy sources as occurring at radiative shocks,\citep{chevalier82,toth93} phase-transition layers,\citep{inutsuka05,inoue06,stone09} ionization-recombination fronts,\citep{williams99,whalen08} or flame and detonation fronts.\citep{short97,daou03} On a cautionary note, I only considered normal perturbations near sharp discontinuities. Taking into account the finite thickness of the transition layer may allow additional instabilities and algebraically growing solutions that the present method is oblivious of.\citep[e.g.,][]{roy86,modestov21}

Other uses of such methods may conceivably be found in computational fluid dynamics and acoustics. For simulation post-processing, one could tag discontinuities and examine their spectral properties a posteriori, or track wave packets in a Monte-Carlo fashion as they are punctually refracted.\citep[e.g.,][]{hertzog02,haviland05} Normal mode decomposition may also help to design boundary conditions\citep{giles88,givoli91} or refine the evaluation of numerical fluxes in conjunction with shock-fitting grid adaptation.\citep[][]{harten83,rawat10} Finally, the accurate characterization of wave propagation can help to design nonintrusive diagnostics for fluid interfaces, as well as means to actively control them and their radiations.

\begin{acknowledgments}
  I am grateful to Pierre Lesaffre for enthusiastically following and proofreading this work and to the MIST team at the Laboratoire de Physique de l'\'Ecole Normale Sup\'erieure where this project started.
  I also thank Florent Renac, Henrik Latter, and Antoine Riols for their constructive comments prior to submission. 
\end{acknowledgments}

\section*{Conflict of Interest Statement}

The author has no conflicts to disclose.

\section*{Data Availability Statement}

The data that support the findings of this study are available from the corresponding author upon reasonable request.

\appendix

\section{Worked example} \label{app:example}

I repeat here the detailed analysis of tangential velocity discontinuities within the proposed framework. I omit magnetic fields and assume that the gas is isothermal and that the perturbations are planar. The general method may seem overcomplicated in this case, but it allows a straightforward extension to more complex and less intuitive cases.

The vector of conservative variables is $\bm{Q}=(\rho,m_x,m_z)$, with $\bm{m}=\rho\bm{V}$. The associated fluxes are $\bm{F}_x = (m_x, m_x^2/\rho + \rho c_s^2, m_x m_z / \rho)$ and $\bm{F}_z = (m_z, m_x m_z/\rho, m_z^2/\rho + \rho c_s^2)$. Linear perturbations satisfy Eq. \eqref{eqn:lindq_dt} away from the interface. Injecting normal modes and taking $\overline{m}_z=0$ in this case yields
\begin{align} \label{eqn:eig}
  \begin{split}
    \left[
    k_x \underbrace{\left(
    \begin{tabular}{ccc}
      $0$ & $1$ & $0$\\
      $c_s^2 - \overline{V}_x^2$ & $2 \overline{V}_x$ & $0$\\
      $0$ & $0$ & $\overline{V}_x$
    \end{tabular}
    \right)}_{\partial \bm{F}_x / \partial \bm{Q}\rvert_{\overline{\bm{Q}}}}
    + k_z \underbrace{\left(
    \begin{tabular}{ccc}
      $0$ & $0$ & $1$\\
      $0$ & $0$ & $\overline{V}_x$\\
      $c_s^2$ & $0$ & $0$
    \end{tabular}
    \right)}_{\partial \bm{F}_z / \partial \bm{Q}\rvert_{\overline{\bm{Q}}}}
    - \omega \Id\right]\cdot \first{\bm{Q}} = 0.
  \end{split}
\end{align}

Let us drop the overline in the following for clarity. The determinant of this system is the dispersion relation
\begin{equation}
  \left(k_x V_x - \omega \right) \left( \omega^2 - 2 k_x V_x \omega + k_x \left[V_x^2 - c_s^2 \right] - k_z^2 c_s^2 \right) = 0.
\end{equation}
Given real $\omega$ and $k_x$, the roots of the dispersion relation are $k_z c_s  = \pm \sqrt{\omega^2 -2k_x V_x \omega + k_x^2 \left(V_x^2 - c_s^2\right)}$. Injecting these wave numbers into Eq. \eqref{eqn:eig} yields two eigenmodes $\first{\bm{Q}} \propto (1, V_x \pm c_s k_x/k_t, \pm c_s k_z/k_t )^T$, where $k_t^2 = k_x^2 + k_z^2$. These correspond to two sound waves that carry the first-order flux perturbations $\first{\bm{F}}_{z} = (\partial \bm{F}_z / \partial \bm{Q})\rvert_{\overline{\bm{Q}}} \cdot \first{\bm{Q}} \propto (\pm c_s k_z / k_t, \pm c_s V_x k_z / k_t,c_s^2)^T$. 

Next, the compatibility conditions between the flux perturbations on both sides of the interface are given by Eq. \eqref{eqn:compat}:
\begin{equation}
  \underbrace{\left(
  \begin{tabular}{ccccccc}
    $-1$ & $0$ & $0$ & $1$ & $0$ & $0$ & $-a_{\rho}$\\
    $0$ & $-1$ & $0$ & $0$ & $1$ & $0$ & $-a_{m_x}$\\
    $0$ & $0$ & $-1$ & $0$ & $0$ & $1$ & $-a_{m_z}$
  \end{tabular}
  \right)}_{\mathcal{H}}
  \cdot
  \left(
  \begin{tabular}{c}
    $\first{\bm{F}}_z^-$\\
    $\first{\bm{F}}_z^+$\\
    $\first{\zeta}$
  \end{tabular}
  \right)
  = 0,
\end{equation}
where $a_q = i (\omega \brac{\overline{q}} -k_x \brac{\overline{f}_{q,x}})$. Solutions are in the nullspace of $\mathcal{H}$, spanned by the following column vectors:
\begin{equation} \label{eqn:null}
  \ker\left(\mathcal{H}\right) =  
  \spn \left\lbrace
  \begin{tabular}{cccc}
    $1$ & $0$ & $0$ & $0$\\
    $0$ & $1$ & $0$ & $0$\\
    $0$ & $0$ & $1$ & $0$\\
    $1$ & $0$ & $0$ & $a_{\rho}$\\
    $0$ & $1$ & $0$ & $a_{m_x}$\\
    $0$ & $0$ & $1$ & $a_{m_z}$\\
    $0$ & $0$ & $0$ & $1$
  \end{tabular}
  \right\rbrace.
\end{equation}
One can clearly discard the (last) row corresponding to $\first{\zeta}$ without affecting the dimension of this nullspace.

Let $\bm{h}_i$ denote the four vectors appearing in Eq. \eqref{eqn:null} but deprived of their $\first{\zeta}$ component. Let $\bm{\ell}_i = (\first{\bm{F}}_{z,i}^-,\bm{0})$ denote the two sound waves living on the left side of the interface, and $\bm{r}_i = (\bm{0},\first{\bm{F}}_{z,i}^+)$ those two living on the right side. One can combine them into the matrix $\mathcal{M} = (\bm{\ell}_1,\bm{\ell}_2,\bm{r}_1,\bm{r}_2,\bm{h}_1,\bm{h}_2,\bm{h}_3,\bm{h}_4)$. This matrix has dimensions $6 \times 8$ and its nullspace is two-dimensional. Any vector in its nullspace describes a superposition of two independent reflection-transmission problems, each having a single incident wave as encoded in its first four components.

\section{Grid-based MHD simulations} \label{app:dns}

For illustrative purposes, I ran a series of grid-based MHD simulations of parallel slow shocks with Ohmic resistivity. One should keep in mind that the phenomena of interest happen near a discontinuity, where the spectral properties of the chosen discretization scheme become critical. On the one hand, shocks are smoothed over a few grid cells and all perturbations are similarly spread in its vicinity. On the other hand, high-order spatial schemes are known to let numerical instabilities attack shocks via the so-called carbuncle phenomenon.\citep[][]{liou2000,robinet2000,kitamura12} Some quantitative disagreement with my predictions is, therefore, expected. 

\subsection{Method}

I used the finite-volume code Pluto \citep{mignone07} version 4.4 on the Cartesian (rectangular) domain $(x,z) \in [ -\uppi, \uppi] \times [-4\uppi, 4\uppi ]$. The initial conditions consisted of an isothermal shock at $z=0$, with upstream Mach number $V_z^-/c_s=2$ and Alfv\'en number $V_z^- / V_A^- = 1/2$, plus a white noise on $(V_x,V_z)$ with amplitude $10^{-6} c_s$. The boundary conditions were periodic in the $x$ (tangential) dimension, while I enforced the initial conditions at the $z$ boundaries. I used a third-order Runge-Kutta time stepping with Courant-Friedrichs-Lewy coefficient $0.4$, a linear reconstruction of primitive variables with the slope limiter of Van Leer,\citep{vanleer74} and the contact-based reconstruction scheme of \citet{gardiner05} for the magnetic field in the constrained transport formalism.\citep{evans88} As for interface fluxes, I used the rather diffusive Lax-Friedrichs scheme to avoid artificial instabilities and also help the corrugation modes stand out of other fluctuations; I consequently used fine grid elements. The mesh was composed of $N_x^2$ square elements over $(x,z) \in [-\uppi,\uppi]^2$, and $N_x \times N_x/2$ geometrically stretched elements near both $z$ boundaries. I ran simulations with $N_x$ ranging from $50$ to $800$ cells in the tangential dimension. Ohmic resistivity was integrated in an operator-split fashion using the super time-stepping scheme described by Alexiades et al.\cite{alexiades96}

\subsection{Results}

\begin{figure}
  \includegraphics[width=\linewidth]{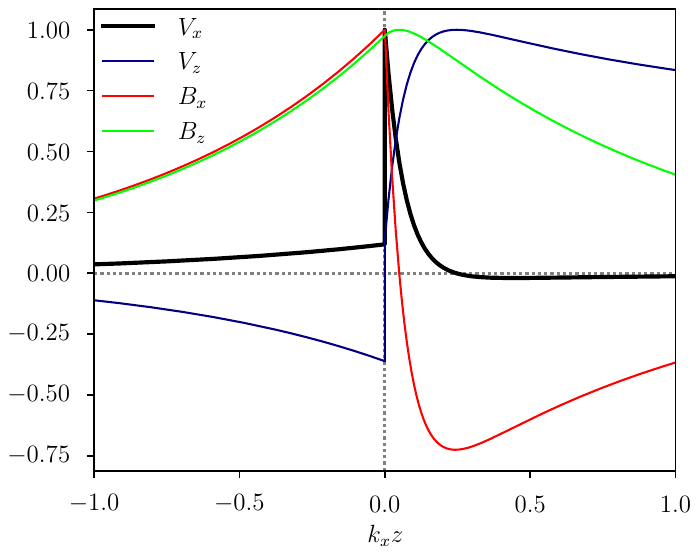}
  \caption{\label{fig:eigenmodes} Predicted structure of the corrugation mode across a slow parallel shock in resistive MHD. The shock is taken to be isothermal with upstream Mach number $V_z^- / c_s = 2$ and resistivity $\eta k_x / V_z^- = 1/4$. The curves were phase-shifted and rescaled for clarity.}
\end{figure}

I show in Fig. \ref{fig:eigenmodes} the predicted structure of the growing modes in the streamwise dimension for $\eta k_x / V_z^- = 1/4$, that is, at the knee of the corresponding curve in Fig. \ref{fig:ohmshock}. While some of the interacting modes slowly decay away from $z=0$, the tangential velocity perturbation features a sharp drop with $k_z \sim 15 k_x$ in magnitude. The $N_x=100$ grid, thus, has roughly $1$ cell per $1/k_z$ characteristic length. It becomes apparent that, even near its resistive stability threshold, accurately resolving the corrugation instability is computationally demanding. 

\begin{figure}
  \includegraphics[width=\linewidth]{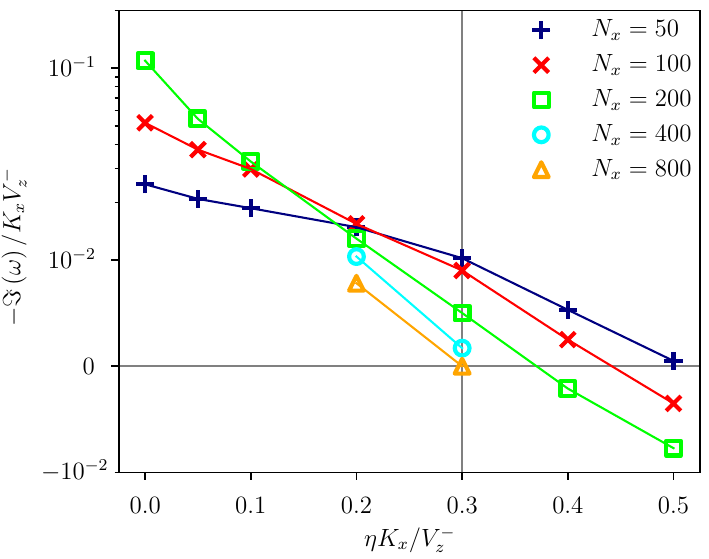}
  \caption{\label{fig:dnsomg} Corrugation growth rate as a function of Ohmic resistivity measured in simulations with different grid resolutions. Note that the stability threshold decreases as grid resolution is increased; the one predicted from Fig. \ref{fig:ohmshock} is marked with a vertical line.}
\end{figure}

Let $K_x=1$ be the smallest tangential wave number resolved on the computational domain. For each grid resolution, I ran a series of simulations with different Ohmic resistivities over a total time of $100/K_x c_s$. All simulations eventually featured a phase of steady exponential growth or decay of their fluctuations. I measured the corresponding rates to better than $10\%$ accuracy and gathered them in Fig. \ref{fig:dnsomg}.

The growth rate decreases with resistivity regardless of grid resolution, as expected from Fig. \ref{fig:ohmshock}. At small resistivity ($\eta K_x / V_z^- < 0.2$), the growth rate increases with the number of grid elements. This is due to small-scale modes being progressively resolved, whose growth rate $\sim k_x V_z^-$ is correspondingly larger (see Fig \ref{fig:SE95}). On the contrary, when resistivity is large enough to stabilize small scales, the exponential growth rates decrease with increasing grid resolution. At the lowest resolution of $N_x=50$, exponential growth is observed up to a resistivity of $\eta K_x / V_z^- = 0.5$. At $N_x=100$, this case is stable but $\eta K_x / V_z^- = 0.4$ is unstable. At $N_x=200$, both previous cases are stable. 

Based on Fig. \ref{fig:ohmshock}, I predict a marginal stability threshold near $0.3$ in this scenario. This threshold is compatible with the trend observed in Fig. \ref{fig:dnsomg} and may indicate that significant grid refinement is required to reach satisfactory convergence. As expected, I did witness the sensitivity of these results when trying different discretization scheme. The fact that growth rates are enhanced at lower resolutions remains intriguing, but its explanation is beyond the goals of this appendix.

\nocite{*}
\bibliography{biblio}

\end{document}